\documentclass{llncs}  

\usepackage{graphicx} 
\usepackage[T1]{fontenc}
\usepackage{amsmath}
\usepackage[english]{babel}
\usepackage{array,tabularx}
\usepackage{booktabs} 
\usepackage{makeidx}  %
\usepackage{xparse}
\usepackage{moredefs}
\usepackage{lips}
\usepackage{epstopdf} 
\usepackage{color}
\usepackage{csquotes}
\usepackage{xcolor}
\usepackage{times}
\usepackage[hidelinks]{hyperref}
\usepackage[nolist,nohyperlinks]{acronym}
\usepackage{comment}
\usepackage{paralist}
\usepackage{cleveref}
\usepackage{wrapfig}
\usepackage{multirow}
\usepackage{microtype}
\usepackage{listings}
\usepackage[super]{nth}
\usepackage{natbib}

\setcitestyle{aysep={,}} 
\usepackage[toc,page]{appendix}
\usepackage{pdflscape}
\usepackage{subcaption}

\usepackage{enumitem}

\usepackage{float}
\floatstyle{plaintop}
\restylefloat{table}

\usepackage[style=base,labelfont=bf,labelsep=period,tableposition=top,font=small]{caption}
\makeatletter
\renewcommand\@biblabel[1]{#1.}
\makeatother
\addto\captionsenglish{}

\usepackage{fancyhdr}

\crefformat{footnote}{#2\footnotemark[#1]#3}
\expandafter\def\expandafter\UrlBreaks\expandafter{\UrlBreaks%
  \do\a\do\b\do\c\do\d\do\e\do\f\do\g\do\h\do\i\do\j%
  \do\k\do\l\do\m\do\n\do\o\do\p\do\q\do\r\do\s\do\t%
  \do\u\do\v\do\w\do\x\do\y\do\z\do\A\do\B\do\C\do\D%
  \do\E\do\F\do\G\do\H\do\I\do\J\do\K\do\L\do\M\do\N%
  \do\O\do\P\do\Q\do\R\do\S\do\T\do\U\do\V\do\W\do\X%
  \do\Y\do\Z}

\newcolumntype{L}[1]{>{\raggedright\arraybackslash}p{#1}}   %
\newcolumntype{C}[1]{>{\centering\arraybackslash}p{#1}}     %
\newcolumntype{R}[1]{>{\raggedleft\arraybackslash}p{#1}}    %

\begin{document}
\frontmatter          %

\mainmatter              %

\title{Designing a Collaborative Platform for Advancing Supply Chain Transparency}

\subtitle{Research Paper} %

\author{Lukas Hueller\inst{1} \and
Tim Kuffner\inst{1} \and
Matthias Schneider\inst{1} \and
Leo Schuhmann\inst{1} \and
Virginie Cauderay\inst{2} \and
Tolga Buz\inst{2} \and
Vincent Beermann\inst{2} \and
Falk Uebernickel\inst{2}}

\authorrunning{Hueller et al.} %

\institute{Hasso Plattner Institute, University of Potsdam, Germany\\
\inst{1} \email{\{firstname.lastname\}@student.hpi.uni-potsdam.de} \\
\inst{2} \email{\{firstname.lastname\}@hpi.de}}

\maketitle
\setcounter{footnote}{0}

\begin{abstract}
Enabling supply chain transparency (SCT) is essential for regulatory compliance and meeting sustainability standards. 
Multi-tier SCT plays a pivotal role in identifying and mitigating an organization's operational, environmental, and social (ESG) risks. 
While research observes increasing efforts towards SCT, a minority of companies are currently publishing supply chain information. 
Using the Design Science Research approach, we develop a collaborative platform for supply chain transparency.
We derive design requirements, formulate design principles, and evaluate the artefact with industry experts.
Our artefact is initialized with publicly available supply chain data through an automated pipeline designed to onboard future participants to our platform.
This work contributes to SCT research by providing insights into the challenges and opportunities of implementing multi-tier SCT and offers a practical solution that encourages organizations to participate in a transparent ecosystem.

\end{abstract}

\section{Introduction}
\label{sec:introduction}

Supply Chain Transparency (SCT) refers to the organizational practice of disclosing detailed information about products and operations \citep{montecchi_supply_2021} and is considered central to regulatory compliance and meeting sustainability standards \citep{montecchi_supply_2021, bai_supply_2020}. Supply chains have grown in complexity over the last decades, with some companies working with thousands of suppliers spread across the globe \citep{mcgrath2021tools}.
SCT thus plays a pivotal role in enabling external stakeholders, such as consumers or investors, to access information concerning an organization's various environmental and social risks \citep{sodhi_research_2019, wilhelm_implementing_2016}. 
For example, disclosure of supply chain information has been shown to influence customers' purchasing behaviors \citep{francisco_supply_2018}, by increasing their willingness to pay \citep{kraft2018supply}.
Additionally, research has identified some of the organizational benefits of acquiring visibility over supply chains for internal stakeholders, such as managers or customers. 
For example, such information allows for better management of the supply chain risks and potential reputational damage or the facilitation of improvements in operational efficiency \citep{sodhi_research_2019}. 

While global companies such as Nike and Adidas have been working on improving their SCT, \citep{searcy2022transformational}, our results show the disclosure of supplier information to remain limited within the sampled companies.
Availability of supply chain data is considered one of the main challenges in this area \citep{krasikov_introducing_2023}.
Reasons for organizational reluctance to disclose supplier information include the fear of losing a competitive advantage \citep{hees_enhancing_2024} or adverse rebound effects in future selection processes \citep{zampou2022design}. 
In addition, companies willing to be transparent remain predominantly concerned with the disclosure of their tier-1 suppliers (i.e., direct suppliers), as expanding the focus beyond these presents significant challenges due to most companies having little to no knowledge of their secondary (i.e., their suppliers' suppliers) and tertiary suppliers \citep{bai_supply_2020, searcy2022transformational}. 

Recent research has proposed different methods to improve SCT through technological innovation, such as IoT \citep{li_iot-based_2017}, blockchain \citep{caro_blockchain-based_2018}, or \textit{Digital Product Passports} \citep{hees_enhancing_2024}. 
However, these ideas still await industry adoption and lack practical implementation and testing, due to high efforts for their implementation.
Commercial databases like those provided by Bloomberg or FactSet offer some supplier relation datasets but are not openly available.
Meanwhile, regulatory pressure on SCT appears to increase, with, for example, the UN sustainable development goals including the topic of supply chain management \citep{srivastava2022supply}, or the European Union's discussions to introduce a Supply Chain Act \citep{dw2024eu}.

Situated at the crossroads between information systems (IS) and supply chain management research, our research focuses on designing a collaborative data platform that centralizes multi-tier supply chain information. 
We propose a platform with an accessible UI and the long-term goal of the relevant industry actors directly providing supply chain information transparently.
An automatic data collection pipeline is created to initialize the platform with publicly available supply chain data. 
Building on the established Design Science Research (DSR) approach \citep{peffers_design_2007}, we identify and motivate the problem of our project, derive
design requirements (DR) to address the identified challenges, formulate design principles (DP), and develop and evaluate the artefact prototype. 
Through research, industry interviews, and our implementation, we further identify the practical challenges and opportunities of implementing multi-tier SCT.

\section{Related Work}
Scholars have identified the lack of standardized terminology relating to the disentanglement of multi-tier supply chains, with many terms being used interchangeably \citep{sodhi_research_2019, sunny_supply_2020, montecchi_supply_2021, roy2021contrasting}. 
We consider the following three core concepts relating to the information management of supply chains: %

\begin{description}%
    \item[Supply Chain Transparency:] ``the practice  of disclosing detailed and  accurate information about operations and products, such as their origin  and sourcing,  manufacturing  processes,  costs,  and logistics'' \textit{\citep[p. 2]{montecchi_supply_2021}}
    \item[Supply Chain Visibility:] ``the extent to which actors within the supply chain (SC) have access to the timely and accurate information that they consider to be key or useful  to  their  operations'' \textit{\cite[p. 1]  {somapa2018characterizing}}
    \item[Supply Chain Traceability:] ``the ability to trace the history, application or location of an entity by means of recorded identifications throughout the entire supply chain'' \textit{\cite[p. 1]{bechini_patterns_2008}}
\end{description}

Namely, the notion of \textit{Supply Chain Transparency} has been discussed to cater to the needs of external stakeholders, such as consumers or NGOs, through the disclosure of information on upstream operations and the product and services sold \citep{sodhi_research_2019}. 
\citet{montecchi_supply_2021} identify six primary angles used to research SCT, including technologies, knowledge integration, governance, sustainability, traceability, and resilience. 
However, to enable transparency, organizations need to be aware of their own supply chain structure \citep{sodhi_research_2019} by establishing \textit{Supply Chain Visibility}.
This refers to the organizational process of gathering relevant supply chain information, primarily intended to be used by internal stakeholders \citep{sodhi_research_2019,montecchi_supply_2021}. 
While visibility is understood to relate to the supply chain level, the notion of \textit{Supply Chain Traceability} occurs on the logistical one \citep{roy2021contrasting}. 
Relating to the acquisition of information to ascertain provenance \citep{sodhi_research_2019} on a granular level \citep{sunny_supply_2020}, it is regarded as essential to meet legal and regulatory requirements \citep{montecchi_supply_2021}. 
Ultimately, traceability enables the acquisition of the required granular data on the logistical level to acquire visibility on the supply chain level internally, the latter in turn enabling the ability of a company to, selectively \citep{sodhi_research_2019}, disclose information to its external stakeholders.

Scholars have investigated the potential of different technologies to support the traceability of products and operations, emphasizing the need for research on their practical applications \citep{sunny_supply_2020, richey_jr_global_2016}.
For example, \citet{li_iot-based_2017} propose an IoT-based tracking and tracing platform for prepacked food, designed to be extendable to other domains.
By envisioning a \textit{Digital Product Passport} for the global supply chain of hydrogen, \citet{hees_enhancing_2024} highlight its potential to reduce information asymmetry and increase data quality.
Similarly to \citet{zampou2022design} and \citet{jussen_issues_2024}, they emphasize the reluctance of supply chain partners to share their data and the need for balance between transparency and confidentiality.
Highlighting the potential of blockchain-based systems to maintain anonymity, \citet{francisco_supply_2018} argue that the utilization and success of its application for supply chain traceability to be bounded by its widespread adoption. 
\citet{sunny_supply_2020}'s research reviews the use of blockchain-based traceability in different industries, highlighting its prevalence in both the pharmaceutical and food industries. 
They further argue for its application for tracking and tracing purposes, regardless of the product or service.

\section{Methodology}

Our research combines an academic and practical approach in a multi-disciplinary effort to anchor research in a more practice-oriented and solution-driven perspective \citep{gholami2016information, kotlarsky2023digital}.
Our current project originated from a \textit{Design Thinking} project on Environmental, Social, and Corporate Governance (ESG) data collection for a large German national bank:
{\it how to present ESG key figures (e.g. ratings, CO$_2$ emissions, water consumption) transparently, comprehensively, and reliably for corporate customers?}
The bank was primarily focused on enhancing transparency across multi-tier supply chains but found the available ESG data and ratings to be frustratingly opaque and lacking in detail.

Due to the practical nature of our research, we follow a DSR approach. DSR is an established paradigm in IS research for solving practical problems in organizational environments \citep{peffers_design_2007, hevner_design_2008}.
Its purpose is to produce design knowledge by developing innovative artefacts as effective solutions for business challenges \citep{winter_design_2008, vom_brocke_special_2020}.

Based on \citet{peffers_design_2007}, we use an adapted version of DSR as introduced by \citet{zampou2022design} with the following four stages: \textit{Problem Identification and Motivation} (section \ref{sec:problem}), \textit{Define the Objectives of the Solution} (section \ref{sec:objectives}), \textit{Design and Development} (section \ref{sec:design}), and \textit{Demonstration and Evaluation} (section \ref{sec:demonstration}). 
In this process, we build on related academic work and industry knowledge to identify and motivate our problem and communicate its purpose, scope, and justificatory knowledge. 
To define the solution's objectives, we build on the identified problems and derive nine DR for our solution. 
Next, we report on the design and development of our artefact prototype, guided by five DP we formulate. 
DP are conceptualized as prescriptive guides that facilitate achieving specific goals, representing actionable knowledge \citep{gregor_anatomy_2007}.  
Eventually, we describe the architecture and UI of the artefact in a demonstration and evaluate it. 

\section{Designing a Collaborative Platform for Transparency in Supply Chains and ESG data}

\subsection{Problem Identification and Motivation}\label{sec:problem}

The lack of centralized and standardized data has been highlighted in the literature as the main challenge for allowing SCT \citep{abraham2019longitudinal, krasikov_introducing_2023}. 
While scholars \citep{sunny_supply_2020, li_iot-based_2017, hees_enhancing_2024} have investigated the potential of different technologies to address such issues, industry adoption and scalability have remained challenging due to a lack of practical implementations and testing due to the usually high effort required. 
Third-party data providers offer an alternative source of information.
Commercial databases such as FactSet or Bloomberg offer comprehensive supplier relation datasets, which undergo review and monitoring by dedicated analyst teams \citep{piraveenan_topology_2020}. 
While these datasets are likely to offer valuable insights, they are proprietary solutions that are not openly available.

We aim to contribute to the academic discourse on SCT and provide a practical solution for its adoption across various supply chain levels. We focus on the technological, organizational, and collaborative criteria required for effective implementation. 
In this context, the scope of our research includes developing a platform that (1) provides an accessible and usable tool for enabling supply chain transparency, (2) motivates industry experts to participate and share their supply chain data, and (3) reduces the need for manual data entry by automating the collection of public supply chain data.
We aim to integrate SCT practices into standard business operations, fostering a more sustainable and ethical global marketplace.

\subsection{Define the Objectives of the Solution}\label{sec:objectives}

The main objective is to create a platform that enables stakeholders to make informed decisions about a companies multi-tier supply chain. 
This platform is shaped by DR derived from challenges we primarily identified through related research (see Section \ref{sec:problem}). 
Table \ref{tab:derived_design_requirements} outlines these DR foundational to the platform's development. 
The platform's attractiveness and value will likely increase with the number of participating companies, leveraging the network effect.
To mitigate the chicken-and-egg dilemma, a typical challenge when launching a platform \citep{stummer_platform_2018}, the platform should be initialized with as much data as possible.
For this purpose, the platform must be initiated with publicly available data to motivate user engagement (DR1). 
Using data that is available publicly, such as Apple's supplier list \citep{apple_inc_apple_2022}, we ensure accessibility for both internal and external stakeholders (DR2).

\begin{table}[H]
\centering
\caption{Nine derived design requirements}
\begin{tabularx}{\textwidth}{@{}l>{\raggedright\arraybackslash}p{2.5cm}>{\raggedright\arraybackslash}X>{\raggedright\arraybackslash}X@{}}
\toprule
\textbf{Label} & \textbf{Name} & \textbf{Description} & \textbf{Rationale} \\
\midrule
DR1 & Initial Data & Platform launch with initial data & Motivate initial user engagement \\ %
DR2 & Public Data & Usage of publicly available data only & Ensure accessibility for all stakeholders \\
DR3 & Automation & Automated data collection pipeline & Address the vast and dynamic global company data \\
DR4 & Data Confidence & Confidence score for data points & Estimate reliability of automatically extracted data \\
DR5 & Metadata & Company metadata inclusion & Aid analytics and insight generation \\
DR6 & Data Predictions & Prediction of hidden supply chain links & Enhance transparency and participation \\
DR7 & Collaboration & Facilitation of data contribution and verification & Foster collaborative data improvement \\
DR8 & Nudging & Incentives for company participation & Encourage transparency and data sharing \\
DR9 & Scalability & Adaption to growing data and user base & Ensure the platform scalability \\
\bottomrule
\end{tabularx}
\label{tab:derived_design_requirements}
\end{table}

With over 300 million companies worldwide as of 2021 \citep{dyvik_estimated_2023} manual data processing is not scalable. %
Extracting supply chain data employing an automated pipeline that runs continuously is one potential solution to this problem (DR3).
However, an automated pipeline requires evaluating the retrieved data's reliability. 
Our core reliability criteria relate to the source of publication of the retrieved data, which could have been published by either the company or a third party. We thus considered the retrieved data to be of higher reliability if we could extract the company's name in both the URL and the text of the data. 
Otherwise, if the company's name could not be found in the URL, only in the text content, we assumed it to be published by a third party, thus of lower reliability.
This evaluation is translated to a confidence score for every data point, estimating its reliability (DR4). 
In addition, the platform should provide useful visualizations regarding key variables (e.g., a transparency rating for a given company, or sustainability-related scores once sufficient data is available). 
To that end, each company in the database is characterized by metadata, e.g., industry, location, revenue, or employee count (DR5). 
To handle missing supply chain data, %
predictions of potential supplier relations can be leveraged, similarly to datasets such as \textit{Bloomberg Global Supply Chain Data} \citep{bloomberg_bloomberg_2024}.
We leverage the metadata collected in DR5 to build a predictive model that estimates supplier relations (DR6).
The platform should allow users to explore and analyze the data, manually manage suppliers, and review predicted relations. 
However, manual adjustments and approvals are reserved for verified company representatives, necessitating robust authentication mechanisms to prevent incorrect data input. 
The general lack of supply chain data may be solved with a collaborative approach \citep{abraham2019longitudinal}, where verified company representatives can improve the data quality of the platform (DR7).
The artefact should motivate companies to participate in the platform and achieve SCT (DR8).
Such motivation can, for example, be achieved through nudges (e.g., using dynamic social norms, such as, \textit{``34\% of companies similar to yours are now sharing their supply chain data''}), and low entry barriers to the platform. 
Lastly, it is important that the platform scales with increasing user requests, data, and network traffic (DR9).

\subsection{Design and Development}\label{sec:design}

\subsubsection{Design Principles for a Collaborative Platform for Transparency in Supply Chains and ESG data.}
As summarised in Table \ref{tab:extended_design_requirements}, we derive five iteratively refined DP to guide the design and development of our platform. These are based on our insights from practically implementing the platform and pipeline, as well as from related work.
Intended to clarify the reasoning behind our design choices, these principles help to generalize our findings and DR, offering a foundational starting point for implementation and benefiting future designers.

Two requirements are central to guarantee data platforms' long-term success, leading to DP1 "Centralization and structure". 
Firstly, the platform must centralize the vastly distributed supply chain data of companies \citep{otto_designing_2019} and must allow the integration of existing technologies \citep[R1.12]{otto_designing_2019}. This is especially important to make supply chains and respective ESG data more transparent.
Secondly, it must standardize the data structure of supplier relations, for example, similar to the GRI standard \citep{gri_global_2024}.

Since users are more likely to join a platform that is already populated with initial data (DP2), our platform includes an automated pipeline to initialize with publicly-available supply chain data (see Section \ref{sec:data_collection_pipeline}). %
As \citet{otto_designing_2019} states in R1.06 of their work, such a platform must be able to first handle and then integrate heterogeneous data into a unified structure.
The latter is required to enable various types of supply chain analysis. %
To further increase data quality, names belonging to the same real-world entity (e.g., company names) must be mapped to each other (DP3). 
Known in research as \textit{entity matching}, such a process brings the challenge of establishing a clear threshold for determining the extent to which an entity should be considered as a match \citep{wang_entity_2011}. 
Our proof of concept implementation addresses this principle with fuzzy matching (i.e., matching "apple" to "apple inc.").
Due to the limited amount of publicly available supply chain data, leveraging data collaboration is necessary to further populate the platform (DP4).
Adding a new company to the graph should be as easy as possible, following the guidance of related work \citep{otto_designing_2019, hees_enhancing_2024, wadhwa_effects_2010, ruijer_designing_2021}.
We, therefore, implement a file upload with which the companies can comfortably add their supplier lists, from which the system automatically extracts supplier names.
As \citet{marzi_b2b_2023} found out, the adoption of such platforms can be quite slow, which is why potential entry barriers have to be reduced and avoided.
To further increase participation, the platform can strategically nudge companies to participate by initially predicting suppliers for missing company supply chains until verified data is provided. 
When an organization adds a supplier, the supplier should be notified. 
This dynamic of indirectly getting pulled onto the platform can serve as an efficient push towards participation (DP5).

\begin{table}[h!]
\centering
\caption{Five formulated design principles}
\begin{tabularx}{\textwidth}{@{}l>{\raggedright\arraybackslash}X>{\raggedright\arraybackslash}X>{\raggedright\arraybackslash}p{1.5cm}>{\raggedright\arraybackslash}X>{\raggedright\arraybackslash}X@{}}
\toprule
\textbf{Label} & \textbf{Design Principle} & \textbf{Aim} & \textbf{Basis DR} & \textbf{Instantiation} & \textbf{Rationale} \\
\midrule
DP1 & Centralization and structure & Serve as gold standard regarding structure and centralization for the data & DR2, DR7 & Centralized platform, with structured data processing and analytic capabilities & Bundling of n:m data distribution to centralization.\\
DP2 & Initial data & Underlying kick-off data to provide value & DR1-DR5, DR8 & Automatic pipeline collecting, harmonizing, and integrating data into the platform & Chicken-and-egg dilemma\\
DP3 & Entity matching & Handle real-world entity matching & DR3, DR7, DR9 & Threshold fuzzy matching of company names & Learnings from implementation; Data heterogeneity must be supported\\
DP4 & Data Collaboratives & Enable data inputs and modifications; Collect and provide data of participating companies & DR7, DR8 & Low-barrier platform UI; handling of file inputs and manual modifications & Onboarding for new companies to the platform must be easy\\
DP5 & Drive participation & Motivate participation in the platform for the higher course of sustainability and overall resilience & DR6, DR8 & Initially predicted suppliers; simple UI; motivation by being part of something bigger; snowball-effect & Mitigate perceived barriers\\
\bottomrule
\end{tabularx}
\label{tab:extended_design_requirements}
\end{table}\textbf{}

\subsubsection{Supplier Relations Data Collection Pipeline.} \label{sec:data_collection_pipeline}
\begin{figure}[h]
    \centering
    \includegraphics[width=1.00\textwidth]{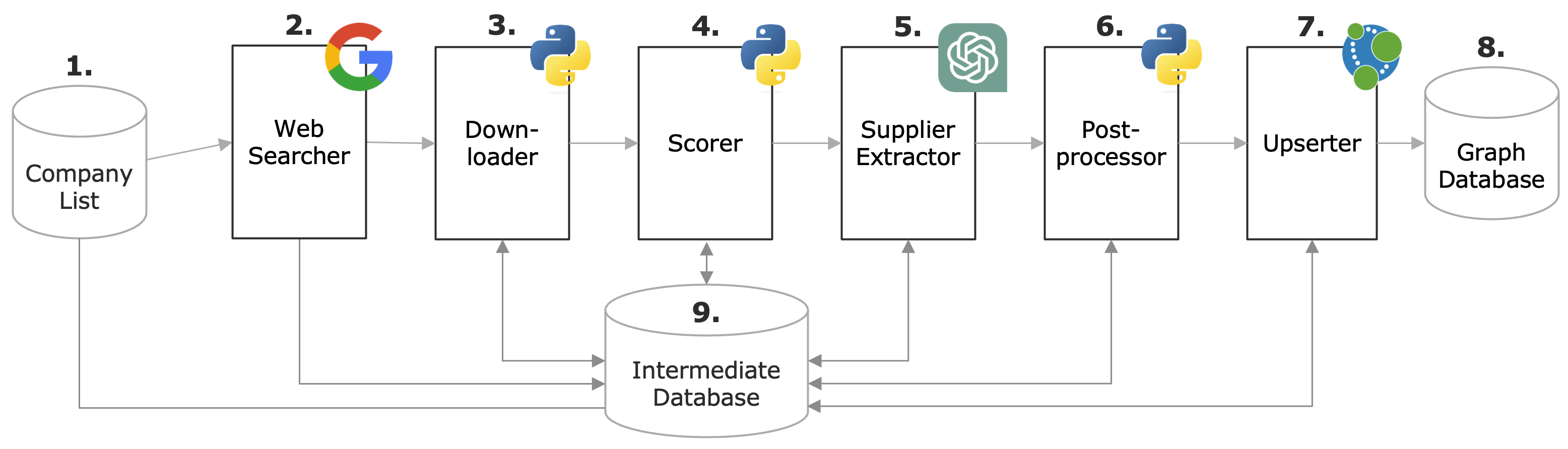}
    \caption{Microservice architecture of the implemented pipeline}
    \label{fig:microservice}
\end{figure}

To answer the chicken-and-egg dilemma (DR1), we propose 
an automated pipeline (DR2). As shown on Figure \ref{fig:microservice}, it consists of nine different microservices (DR3). The starting point of the pipeline is a list of known companies (1 in Figure \ref{fig:microservice}). For our prototype, we select all 5,679 companies with a market capitalization of at least USD \$1 billion, retrieved from Marketscreener.com (as of January 10th, 2024).
Thereupon, a web searcher (2) returning potential supplier disclosure documents for the respective company is implemented. 
The pipeline then downloads (3) the files and extracts the required data.

Next, the extracted data is evaluated using a scoring function (4). 
This function considers the extracted file text and its source (URL) to determine if the document is reliable (DR4).
To extract companies' names from the text, we use OpenAI's GPT-4 API as an entity recognizer (5). 
To mitigate GPT-4 hallucinations (i.e., false positives), we validate the identified entities by matching them against the list of 5,679 known companies (as mentioned in (1)), using fuzzy matching during post-processing (6).
This approach does not consider suppliers outside the initial set, but it significantly reduces false positives and improves data quality. 
Finally, we upsert (7) the identified supplier relations into our graph database (8).
We utilize an intermediate database (9) to share data among the different microservices (2-7) during the process.

As stated in DR6, predictions of supplier relations are required due to the limited availability of public supply chain data. %
We thus enrich each company from the database with metadata attributes from Wikidata (DR5), which are used as input for our rule-based prediction algorithm. 
The algorithm ranks the companies with the most supplier relations for each industry (e.g., Information Technology and Software) and region (e.g., North America). 
This suggests that the highest-ranked companies are potential suppliers for companies within the same industry and region combination.

The initial graph constructed by the automated pipeline contains the 5,679 sampled companies, for 1,396 (i.e. 24\%) of which supplier relations are found (without predictions). 
2,467 relations are identified by sourcing information from 762 files classified as reliable (i.e., a confidence score greater than 0.6 with a max of 1.0). 
Our results highlight that a majority of companies currently do not share their supplier information, emphasising the necessity of our platform. %

\subsubsection{Supplier Relation Platform and UI.}
\begin{figure}[htp]
    \centering
    \includegraphics[width=1\textwidth]{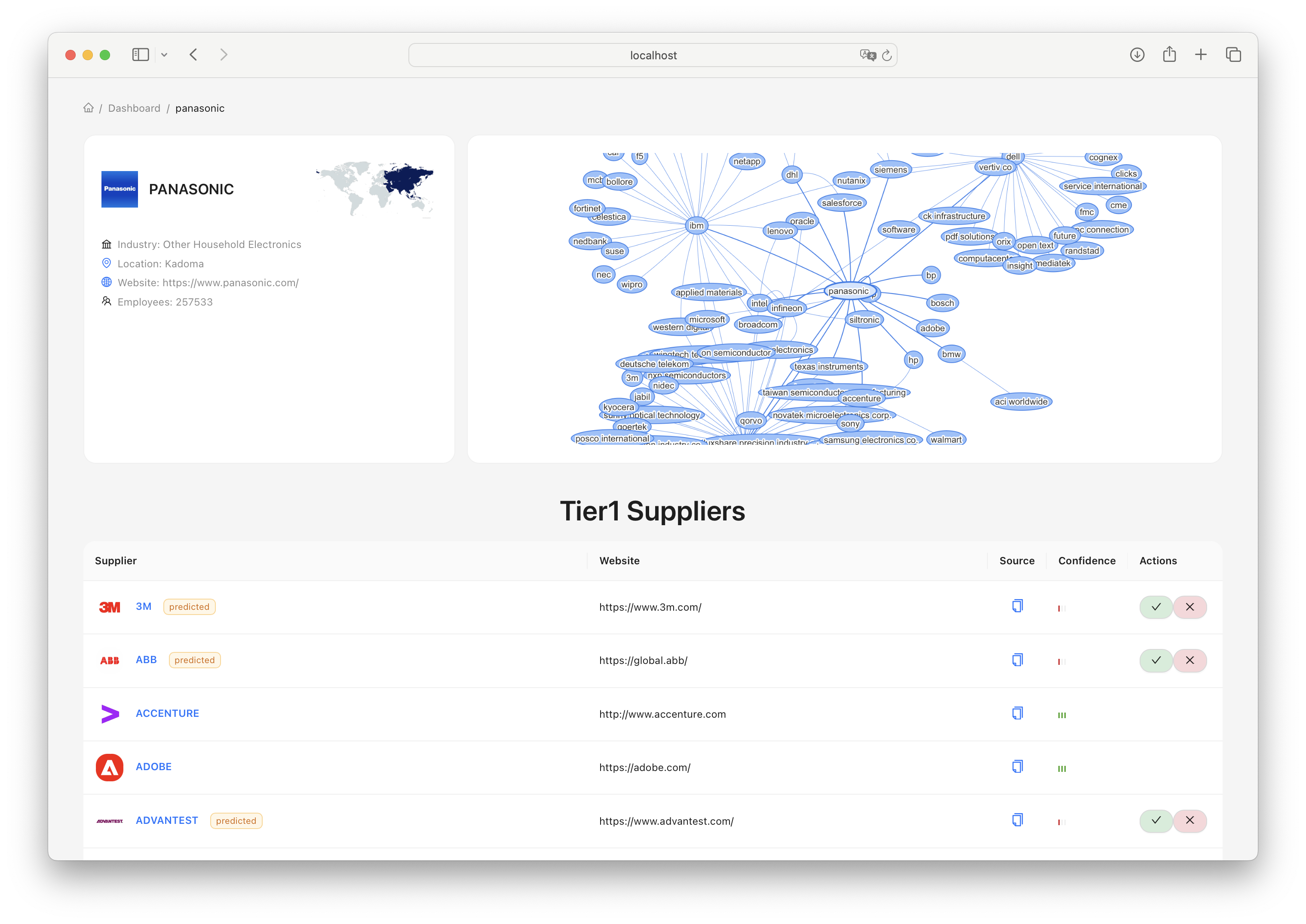}
    \caption{Supplier relation platform user interface with network graph}
    \label{fig:frontend}
\end{figure}

We build a starting point for a scalable platform (DR9), enabling transparency and data analysis of supply chains based on an initial dataset (DR7). 
The React front end (see Figure \ref{fig:frontend}) displays the identified and predicted supplier relations between companies through a graph and table, enriched with the data source and the confidence score (DR4).

All supplier relations are subject to confirmation or rejection by authenticated company representatives. 
Moreover, authorized users possess the capability to incorporate suppliers either manually or through file upload. 
Subsequently, newly added companies receive an email notification (if the email is known, e.g., from Wikidata) informing them of their inclusion as a supplier to a specific company. 
This dynamic has the potential to catalyze a network effect (DR8).

\subsection{Demonstration and Evaluation}\label{sec:demonstration}
Firstly, we %
review the quality and completeness of the data collected by our automated pipeline. %
Secondly, we evaluate the DR and principles with expert interviews to validate the resulting artefact.

\subsubsection{Evaluation of the Supplier Relation Dataset.}

\emph{How well has our automatic pipeline identified documentation relating to the supply chain and the suppliers contained within it?}
To validate the completeness of our dataset, we manually reviewed the available supply chain data for the ten largest stock-listed companies across North America, Europe, and Asia (i.e., 30 companies). 
As shown in Table \ref{tab:quantitative_eva}, out of the documents found for these 30 companies, our pipeline had adequately identified the seven companies (i.e., 23\%), that have published supplier lists. 
Our pipeline extracts 883 of the 1,391 (63.5\%) supplier companies manually identified from those supplier lists. 
In summary, our pipeline identifies all companies from this subset that publish their supply chain data and most of their suppliers.
However, our fuzzy matching yields a much smaller number of companies due to our limited list of known companies explained above.
Extending the list of known companies in future work and using additional matching techniques will likely improve accuracy.

\begin{table}[ht]
\centering
\caption{Quantitative evaluation of top 10 companies per region by market capitalization, showing that our automated pipeline is accurate in identifying published supplier lists. Detecting all mentioned suppliers is a more difficult task.}
\label{tab:quantitative_eva}
\begin{tabular}{lcccccc}
\toprule
\textbf{Region} & \textbf{Companies} & \textbf{Published} & \textbf{Identified} & \textbf{Suppliers} & \textbf{Suppliers} & \textbf{Matched} \\
 & \textbf{probed} & \textbf{lists} & \textbf{lists} & \textbf{across lists} & \textbf{across lists} & \textbf{Initial list} \\
 &  & \textbf{(checked)} & \textbf{(auto)} & \textbf{(checked)} & \textbf{(auto)} & \textbf{(auto)} \\
\midrule
North America           & 10 & 2 & 2 & 639 & 303 & 62 \\
Europe                  & 10 & 2 & 2 & 125 & 122 & 2 \\
Asia and Middle East    & 10 & 3 & 3 & 627 & 458 & 31 \\
\midrule
\textbf{Total}          & \textbf{30} & \textbf{7} & \textbf{7} & \textbf{1391} & \textbf{883} & \textbf{95} \\
\bottomrule
\end{tabular}
\end{table}

\emph{Which insights can we extract from our supply chain database?}
For this analysis, we define a simplified proxy for a company's transparency. Specifically, we consider a company to be transparent if our pipeline is able to extract at least one supplier from public data. This suggests that the company discloses supplier information, thereby achieving a certain level of transparency.
Our results show that transparency is equally distributed across all industries, although there are differences between regions. %
As shown in Table \ref{tab:distr_of_transp}, organizations from Europe and North America appear more transparent, with respectively 14.84\% and 9.67\% of the evaluated companies from these regions disclosing suppliers. 
The transparency percentage of Africa lies at similar heights, at 10.71\%. 
However, the reliability of this measure is lower due to the smaller-sized sample of companies included in our list. %
On average, 7.24\% of the sampled companies from the remaining continents (i.e., Asia, Middle East, Oceania, South America) disclose suppliers. 
Our results show a global average of 10.32\% of companies publishing supply chain data across the sample companies.

\begin{table}[ht]
\centering
\caption{Distribution of transparency across continents (i.e., companies that publish their supplier lists)}
\label{tab:transparent_companies}
\begin{tabular}{lccc}
\toprule
\textbf{Continent} & \textbf{Evaluated Companies} & \textbf{Transparent Companies} & \textbf{Transparency (\%)} \\
\midrule
Europe (EU)        & 1,435   & 213   & 14.84\% \\
Africa (AF)        & 84      & 9     & 10.71\%  \\
North America (NA) & 2,564   & 248   & 9.67\% \\
Asia (AS)          & 790     & 67    & 8.48\%  \\
South America (SA) & 225     & 19    & 8.44\%  \\
Oceania (OC)       & 209     & 16    & 7.66\%  \\
Middle East (ME)   & 364     & 13    & 3.57\%  \\
\midrule
\textbf{Total} & \textbf{5,671}  & \textbf{585}  & \textbf{10.32\%} \\
\bottomrule
\end{tabular}
\label{tab:distr_of_transp}
\end{table}

\subsubsection{Expert Evaluation and Feedback.}
To assess the effectiveness of our platform in improving SCT, we conducted five expert interviews with participants from different industries.
We engaged with two sustainability and innovation experts from the banking sector, %
an operational IT procurement manager from the IT services sector, a consulting services procurement professional, and an operational automotive procurement specialist.
The main goal of these interviews was to evaluate the utility of the platform, its ease of use, and its capacity to support organizational goals, such as enhancing operational efficiency or improving supplier engagement. 
Each interviewee was granted access to the platform and encouraged to independently explore its features, without direct guidance from our research team. 
They were then asked to share their feedback and insights about the platform's practical applications.

Positive feedback centered around the analytical capabilities of the platform, namely its potential to inform critical decision-making processes based on specific criteria such as supplier reliability and sustainability/ESG metrics, and its cost efficiency, as it functions without hardware investments or complex data integration processes.  
Additionally, several interviewees acknowledged that their companies had already conducted manual due diligence on their suppliers, confirming the relevance of adopting such automated solutions: \\

\noindent \textit{``For big general agreements, we would do such background checks.''} \\ -- Procurement expert, IT Services SME \\

\noindent \textit{``A tool that shows all the supplier relations can be useful since we already investigate that.''} -- Procurement expert, Automotive \\

Another positive aspect was the platform's potential to increase stakeholder participation. 
One interviewee explained that such a platform could empower small and medium enterprises to make informed decisions, as these entities currently lack the resources and knowledge to process such information (Sustainability Expert, Banking). 
All interviewed experts were willing to engage with the platform. %
Some indicated a preference for employing the platform primarily for decision-making tasks, while others highlighted the need for such a platform to integrate smoothly with existing supplier management or sustainability tools. 
Regarding the organizational willingness to disclose supplier data, responses varied. 
Some interviewees were inclined to share their suppliers based on transparency principles, while others cited the need for approval from senior decision-makers within their organizations due to confidentiality concerns. 
Further considerations included smaller procurement organizations' potential to seek simpler systems for managing supplier relations and prioritizing essential data for decision-making, such as cost, delivery times, or product quality. 
Moreover, while price and availability remain essential in decision-making, sustainability is often a secondary consideration. 
Additional suggestions included incentivizing more sustainable supplier choices through certifications or integrating the platform with existing systems like Salesforce.

\section{Conclusion and Outlook}

The integration of supplier data into shared information systems is essential for improving transparency, mitigating risks, and advancing sustainability efforts. While some research has been conducted on the practical application of technology for supply chain traceability, less is known from a SCT perspective. 
Our research thus contributes to this endeavor by proposing a practical solution and highlighting the importance of collaborative data sharing in increasing the transparency of the global supply chain landscape.
Similarly to \citet{hees_enhancing_2024}, we aimed to increase transparency by incentivizing cross-organizational data collaboration.
Throughout the design of our platform, our research unveiled five key DP, emphasizing centralization, structure, motivation to participate, the necessity of harmonization, and collaboration capabilities (DP1-DP5). 
We identify these to be crucial for the success of a centralized multi-tier supply chain data platform, aimed at addressing the scarcity of structured data \citep{abraham2019longitudinal, krasikov_introducing_2023} and providing practical solutions \citep{richey_jr_global_2016, recker_communications_2009}.
This is especially pertinent for ensuring regulatory compliance and quality assurance \citep{montecchi_supply_2021} and discovering hidden risks (e.g., unethical practices) across multi-tier supply chains \citep{wilhelm_implementing_2016}.
In addition, both internal and external stakeholders can derive benefits from the insights provided by transparent multi-tier supply chain data.
While internal stakeholders may be incentivised to improve their supplier selection process, external stakeholders can acquire the relevant information to make better-informed choices, e.g., regarding purchased products and services.

Our platform provides a starting point for practical implementation. For future research, we recommend analysing the most fitting business model and operators and further improving the likelihood of adoption. Our implementation also derives DP generally applicable to any cross-organizational data as well as the incorporation of requirements identified by \citet{otto_designing_2019} and \citet{marzi_b2b_2023}.
We build an automated pipeline to initialize the platform with publicly available data to mitigate one important challenge of establishing such platforms. %
In addition, our artifact differentiates from previous work by envisioning a global collaboration platform that is not bound to the adoption of specific technology, such as IoT or blockchain, some of which require hardware investments but can be accessed via a standard web interface. 
Our interview results reveal organizational interest in leveraging the platform for its analytical capabilities. 
However, opinions varied concerning the sharing of supplier information. 
The prevailing sentiment indicates a potential for adoption, especially if the platform remains easy to use and freely accessible, addresses integration issues, and demonstrates clear benefits in decision-making. 
Ultimately, our study contributes research on SCT by extending the current knowledge on the practical application of digital technologies for such endeavors. 
Future research can build upon the knowledge acquired throughout the artefact design, leveraging the challenges and opportunities identified as a future research direction.

\subsubsection{Limitations.} 
It has proven difficult to find industry experts for interviews -- we recommend conducting additional interviews in future work to further validate our insights.
Our list of the largest 5,679 companies provides a starting point for the platform's database, but limits the entity matching algorithm's effectiveness -- many suppliers are smaller companies. %
To further improve the system's capabilities to recognize supplier companies, we recommend 
(1) leveraging a more advanced language model than GPT-4 (once available), (2) implementing a scalable matching algorithm that robustly identifies company variants, and (3) growing the list of known companies as much as possible while ensuring that companies can still be identified distinctly. %
Furthermore, a collaborative data platform for SCT relies on organizational trust without third-party verification -- %
further work should investigate how to enable trust in such an environment.

\bibliographystyle{agsm}
\bibliography{literature}

@article{srivastava2022supply,
  title={Supply Chain Management and the United Nations Sustainable Development Goals},
  author={Srivastava, Ashutosh and Vyas, Vidhisha and Gurtu, Amulya},
  journal={Operations and Supply Chain Management: An International Journal},
  volume={15},
  number={4},
  pages={505--515},
  year={2022}
}

@article{vom_brocke_special_2020,
	title = {Special Issue Editorial – Accumulation and Evolution of Design Knowledge in Design Science Research: A Journey Through Time and Space},
    year = {2020},
	volume = {21},
	issn = {15369323},
	pages = {520--544},
	number = {3},
	journaltitle = {Journal of the Association for Information Systems},
	shortjournal = {{JAIS}},
	author = {Vom Brocke, Jan and Winter, Robert and Hevner, Alan and Maedche, Alexander},
	date = {2020-05-01}
}

@article{sunny_supply_2020,
	title = {Supply chain transparency through blockchain-based traceability: {An} overview with demonstration},
	volume = {150},
	journal = {Computers \& Industrial Engineering},
	author = {Sunny, Justin and Undralla, Naveen and Pillai, V Madhusudanan},
	year = {2020},
	note = {Publisher: Elsevier},
	pages = {106895},
}

@article{wadhwa_effects_2010,
	title = {Effects of information transparency and cooperation on supply chain performance: a simulation study},
	volume = {48},
	number = {1},
	journal = {International Journal of Production Research},
	author = {Wadhwa, Subhash and Mishra, Madhawanand and Chan, Felix TS and Ducq, Yves},
	year = {2010},
	note = {Publisher: Taylor \& Francis},
	pages = {145--166},
}

@article{abraham2019longitudinal,
  title={A longitudinal exploratory investigation of innovation systems and sustainability maturity using case studies in three industries},
  author={Abraham, Thomas and Dao, Viet T},
  journal={Journal of Enterprise Information Management},
  volume={32},
  number={4},
  pages={668--687},
  year={2019},
  publisher={Emerald Publishing Limited}
}

@article{krasikov_introducing_2023,
	title = {Introducing a {Data} {Perspective} to {Sustainability}: {How} {Companies} {Develop} {Data} {Sourcing} {Practices} for {Sustainability} {Initiatives}},
	volume = {53},
	number = {1},
	journal = {Communications of the Association for Information Systems},
	author = {Krasikov, Pavel and Legner, Christine},
	year = {2023},
	pages = {5},
}

@article{richey_jr_global_2016,
	title = {A global exploration of big data in the supply chain},
	volume = {46},
	number = {8},
	journal = {International Journal of Physical Distribution \& Logistics Management},
	author = {Richey Jr, Robert Glenn and Morgan, Tyler R and Lindsey-Hall, Kristina and Adams, Frank G},
	year = {2016},
	note = {Publisher: Emerald Group Publishing Limited},
	pages = {710--739},
}

@article{ruijer_designing_2021,
	title = {Designing and implementing data collaboratives: {A} governance perspective},
	volume = {38},
	issn = {0740-624X},
	url = {https://www.sciencedirect.com/science/article/pii/S0740624X21000484},
	number = {4},
	journal = {Government Information Quarterly},
	author = {Ruijer, Erna},
	year = {2021},
	pages = {101612},
}

@article{jussen_issues_2024,
	title = {Issues in {Inter}-{Organizational} {Data} {Sharing}: {Findings} from {Practice} and {Research} {Challenges}},
	journal = {Data \& Knowledge Engineering},
	author = {Jussen, Ilka and Möller, Frederik and Schweihoff, Julia and Gieß, Anna and Giussani, Giulia and Otto, Boris},
	year = {2024},
	note = {Publisher: Elsevier},
	pages = {102280},
}

@article{sodhi_research_2019,
	title = {Research opportunities in supply chain transparency},
	volume = {28},
	number = {12},
	journal = {Production and Operations Management},
	author = {Sodhi, ManMohan S and Tang, Christopher S},
	year = {2019},
	note = {Publisher: Wiley Online Library},
	pages = {2946--2959},
}

@article{montecchi_supply_2021,
	title = {Supply chain transparency: {A} bibliometric review and research agenda},
	volume = {238},
	journal = {International Journal of Production Economics},
	author = {Montecchi, Matteo and Plangger, Kirk and West, Douglas C},
	year = {2021},
	note = {Publisher: Elsevier},
	pages = {108152},
}

@article{stummer_platform_2018,
	title = {Platform launch strategies},
	volume = {60},
	journal = {Business \& Information Systems Engineering},
	author = {Stummer, Christian and Kundisch, Dennis and Decker, Reinhold},
	year = {2018},
	note = {Publisher: Springer},
	pages = {167--173},
}

@inproceedings{caro_blockchain-based_2018,
	title = {Blockchain-based traceability in {Agri}-{Food} supply chain management: {A} practical implementation},
	booktitle = {2018 {IoT} {Vertical} and {Topical} {Summit} on {Agriculture} - {Tuscany} ({IOT} {Tuscany})},
	author = {Caro, Miguel Pincheira and Ali, Muhammad Salek and Vecchio, Massimo and Giaffreda, Raffaele},
	year = {2018},
	pages = {1--4},
}

@article{wang_entity_2011,
	title = {Entity matching: {How} similar is similar},
	volume = {4},
	number = {10},
	journal = {Proceedings of the VLDB Endowment},
	author = {Wang, Jiannan and Li, Guoliang and Yu, Jeffrey Xu and Feng, Jianhua},
	year = {2011},
	note = {Publisher: VLDB Endowment},
	pages = {622--633},
}

@misc{bloomberg_bloomberg_2024,
	title = {Bloomberg {Global} {Supply} {Chain} {Data}},
	url = {https://www.bloomberg.com/professional/dataset/global-supply-chain-data/},
	author = {{Bloomberg}},
	year = {2024},
}

@misc{dyvik_estimated_2023,
	title = {Estimated number of companies worldwide from 2000 to 2021},
	url = {https://www.statista.com/statistics/1260686/global-companies/},
	urldate = {2024-02-13},
	author = {Dyvik, Einar H.},
	year = {2023},
}

@misc{apple_inc_apple_2022,
	title = {Apple {Supplier} {List}},
	url = {https://www.apple.com/supplier-responsibility/pdf/Apple-Supplier-List.pdf},
	author = {{Apple Inc.}},
	year = {2022},
}

@misc{gri_global_2024,
	title = {Global {Reporting} {Initiative}},
	url = {https://www.globalreporting.org},
	author = {{GRI}},
	year = {2024},
}

@article{bai_supply_2020,
	title = {A supply chain transparency and sustainability technology appraisal model for blockchain technology},
	volume = {58},
	number = {7},
	journal = {International Journal of Production Research},
	author = {Bai, Chunguang and Sarkis, Joseph},
	year = {2020},
	note = {Publisher: Taylor \& Francis},
	pages = {2142--2162},
}

@article{bechini_patterns_2008,
	title = {Patterns and technologies for enabling supply chain traceability through collaborative e-business},
	volume = {50},
	number = {4},
	journal = {Information and software technology},
	author = {Bechini, Alessio and Cimino, Mario GCA and Marcelloni, Francesco and Tomasi, Andrea},
	year = {2008},
	note = {Publisher: Elsevier},
	pages = {342--359},
}

@article{wilhelm_implementing_2016,
	title = {Implementing sustainability in multi-tier supply chains: {Strategies} and contingencies in managing sub-suppliers},
	volume = {182},
	journal = {International Journal of Production Economics},
	author = {Wilhelm, Miriam and Blome, Constantin and Wieck, Ellen and Xiao, Cheng Yong},
	year = {2016},
	note = {Publisher: Elsevier},
	pages = {196--212},
}

@article{francisco_supply_2018,
	title = {The supply chain has no clothes: {Technology} adoption of blockchain for supply chain transparency},
	volume = {2},
	number = {1},
	journal = {Logistics},
	author = {Francisco, Kristoffer and Swanson, David},
	year = {2018},
	note = {Publisher: MDPI},
	pages = {2},
}

@article{li_iot-based_2017,
	title = {{IoT}-based tracking and tracing platform for prepackaged food supply chain},
	volume = {117},
	number = {9},
	journal = {Industrial Management \& Data Systems},
	author = {Li, Zhi and Liu, Guo and Liu, Layne and Lai, Xinjun and Xu, Gangyan},
	year = {2017},
	note = {Publisher: Emerald Publishing Limited},
	pages = {1906--1916},
}

@article{hees_enhancing_2024,
	title = {Enhancing trust in global supply chains: {Conceptualizing} {Digital} {Product} {Passports} for a low-carbon hydrogen market},
	volume = {34},
	number = {1},
	journal = {Electronic Markets},
	author = {Heeß, Paula and Rockstuhl, Jakob and Körner, Marc-Fabian and Strüker, Jens},
	year = {2024},
	note = {Publisher: Springer},
	pages = {1--20},
}

@article{piraveenan_topology_2020,
	title = {Topology of {International} {Supply} {Chain} {Networks}: {A} {Case} {Study} {Using} {Factset} {Revere} {Datasets}},
	volume = {8},
	journal = {IEEE Access},
	author = {Piraveenan, Mahendra and Jing, Hongze and Matous, Petr and Todo, Yasuyuki},
	year = {2020},
	pages = {154540--154559},
}

@article{recker_communications_2009,
	title = {{ACIS 2007 Panel Report: Lack of Relevance in IS Research}},
	volume = {24},
	number = {18},
	journal = {Communications of the Association for Information Systems},
	author = {Recker, Jan and Young, Raymond and Darroch, Fiona and Marshall, Peter and McKay, Judy},
	year = {2009},
	pages = {303--314},
}

@article{otto_designing_2019,
	title = {Designing a multi-sided data platform: findings from the {International} {Data} {Spaces} case},
	volume = {29},
	number = {4},
	journal = {Electronic Markets},
	author = {Otto, Boris and Jarke, Matthias},
	year = {2019},
	note = {Publisher: Springer},
	pages = {561--580},
}

@article{marzi_b2b_2023,
	title = {{B2B} digital platform adoption by {SMEs} and large firms: {Pathways} and pitfalls},
	volume = {114},
	journal = {Industrial Marketing Management},
	author = {Marzi, Giacomo and Marrucci, Anna and Vianelli, Donata and Ciappei, Cristiano},
	year = {2023},
	note = {Publisher: Elsevier},
	pages = {80--93},
}

@article{hevner_design_2008,
	title = {Design science in information systems research},
	volume = {28},
	number = {1},
	journal = {Management Information Systems Quarterly},
	author = {Hevner, Alan R and March, Salvatore T and Park, Jinsoo and Ram, Sudha},
	year = {2008},
	pages = {6},
}

@article{winter_design_2008,
	title = {Design science research in {Europe}},
	volume = {17},
	journal = {European Journal of Information Systems},
	author = {Winter, Robert},
	year = {2008},
	note = {Publisher: Springer},
	pages = {470--475},
}

@inproceedings{gregor_anatomy_2007,
	title = {The anatomy of a design theory},
	publisher = {Association for Information Systems},
	author = {Gregor, Shirley and Jones, David and {others}},
	year = {2007},
}

@article{peffers_design_2007,
	title = {A design science research methodology for information systems research},
	volume = {24},
	number = {3},
	journal = {Journal of management information systems},
	author = {Peffers, Ken and Tuunanen, Tuure and Rothenberger, Marcus A and Chatterjee, Samir},
	year = {2007},
	note = {Publisher: Taylor \& Francis},
	pages = {45--77},
}

@article{kraft2018supply,
  title={Supply chain visibility and social responsibility: Investigating consumers’ behaviors and motives},
  author={Kraft, Tim and Vald{\'e}s, Le{\'o}n and Zheng, Yanchong},
  journal={Manufacturing \& Service Operations Management},
  volume={20},
  number={4},
  pages={617--636},
  year={2018},
  publisher={INFORMS}
}

@article{gholami2016information,
  title={Information systems solutions for environmental sustainability: How can we do more?},
  author={Gholami, Roya and Watson, Richard T and Hasan, Helen and Molla, Alemayehu and Bjorn-Andersen, Niels},
  journal={Journal of the Association for Information Systems},
  volume={17},
  number={8},
  pages={2},
  year={2016}
}

@article{kotlarsky2023digital,
  title={Digital Sustainability in Information Systems Research: Conceptual Foundations and Future Directions},
  author={Kotlarsky, Julia and Oshri, Ilan and Sekulic, Nevena},
  journal={Journal of the Association for Information Systems},
  volume={24},
  number={4},
  pages={936--952},
  year={2023}
}

@article{zampou2022design,
  title={A design theory for energy and carbon management systems in the supply chain},
  author={Zampou, Eleni and Mourtos, Ioannis and Pramatari, Katerina and Seidel, Stefan},
  journal={Journal of the Association for Information Systems},
  volume={23},
  number={1},
  pages={329--371},
  year={2022}
}

@article{roy2021contrasting,
  title={Contrasting supply chain traceability and supply chain visibility: are they interchangeable?},
  author={Roy, Vivek},
  journal={The International Journal of Logistics Management},
  volume={32},
  number={3},
  pages={942--972},
  year={2021},
  publisher={Emerald Publishing Limited}
}

@article{somapa2018characterizing,
  title={Characterizing supply chain visibility--a literature review},
  author={Somapa, Sirirat and Cools, Martine and Dullaert, Wout},
  journal={The International Journal of Logistics Management},
  volume={29},
  number={1},
  pages={308--339},
  year={2018},
  publisher={Emerald Publishing Limited}
}

@article{searcy2022transformational,
  title={Transformational transparency in supply chains: Leveraging technology to drive radical change},
  author={Searcy, Cory and Castka, Pavel and Mohr, Jakki and Fischer, S{\"o}nke},
  journal={California management review},
  volume={65},
  number={1},
  pages={19--43},
  year={2022},
  publisher={SAGE Publications Sage CA: Los Angeles, CA}
}

@article{mcgrath2021tools,
  title={Tools and technologies of transparency in sustainable global supply chains},
  author={McGrath, Paul and McCarthy, Lucy and Marshall, Donna and Rehme, Jakob},
  journal={California Management Review},
  volume={64},
  number={1},
  pages={67--89},
  year={2021},
  publisher={Sage Publications Sage CA: Los Angeles, CA}
}

@misc{dw2024eu,
    title = {{EU countries back new human rights supply chain law}},
    author = {{Deutsche Welle}},
    year = {2024},
    url = {https://www.dw.com/en/eu-countries-back-new-human-rights-supply-chain-law/a-68577888}
}

\end{document}